\documentclass[aps,preprint,showpacs, showkeys, dvips]{revtex4}%
\usepackage{amsfonts}
\usepackage{amsmath}
\usepackage{amssymb}
\usepackage{graphicx}
\usepackage{float}
\usepackage{appendix}%
\providecommand{\U}[1]{\protect \rule{.1in}{.1in}}
\numberwithin{equation}{section}
\begin{document}
\title[ ]{Vacuum Stability of the $\mathcal{PT}$-Symmetric $\left(  -\phi^{4}\right)  $
Scalar Field Theory }
\author{Abouzeid M. Shalaby}
\email{amshalab@mans.edu.eg}
\affiliation{Physics Department, Faculty of Science, Mansoura University, Egypt}

\begin{abstract}
In this work, we study the vacuum stability of the classical unstable $\left(
-\phi^{4}\right)  $ scalar field potential. Regarding this, we obtained the
effective potential, up to second order in the coupling, for the theory in
$1+1$ and $2+1$ space-time dimensions. We found that the obtained effective
potential is bounded from below, which proves the vacuum stability of the
theory in space-time dimensions higher than the previously studied $0+1$ case.
In our calculations, we used the canonical quantization regime in which one
deals with operators rather than classical functions used in the path integral
formulation. Therefore, the non-Hermiticity of the effective field theory is
obvious. Moreover, the method we employ implements the canonical equal-time
commutation relations and the Heisenberg picture for the operators. Thus, the
metric operator is implemented in the calculations of the transition
amplitudes. Accordingly, the method avoids the very complicated calculations
needed in other methods for the metric operator. To test the accuracy of our
results, we obtained the exponential behavior of the vacuum condensate for
small coupling values, which has been obtained in the literature using other
methods. We assert that this work is interesting, as all the studies in the
literature advocate the stability of the $\left(  -\phi^{4}\right)  $ theory
at the quantum mechanical level while our work extends the argument to the
level of field quantization.

\end{abstract}
\maketitle

\section{ Introduction\label{intro}}

Among its wide range of applications, the subject of $\mathcal{PT}$-symmetric
theories has stressed the bounded-from-above $(-x^{4})$ quantum mechanical
potential \cite{bendr,qop,bendrcop,jonespath,jonesqop,aboqm}. The recipe for
the calculations within such theories is to choose a specific contour in the
complex $x$-plane and apply the quantization condition ($\chi_{n}
\rightarrow0$ as $\left \vert x\right \vert \rightarrow \infty$) on the wave
functions $\chi_{n}$. It is this boundary condition that renders the problem
non-Hermitian and $\mathcal{PT}$-symmetric as well. For the complete
determination of the transition amplitudes within the $\mathcal{PT}$-symmetric
theories, the positive definite metric operator and thus the equivalent
Hermitian Hamiltonian have to be obtained. This has been done for the
$\mathcal{PT}$-symmetric ($-x^{4}$) theory in Ref. \cite{jonesqop}.
Remarkably, the equivalent Hermitian Hamiltonian is bounded from below. This
gives no doubt that the spectrum of the $\mathcal{PT}$-symmetric ($-x^{4}$)
theory is stable. However, in higher space-time dimensions \textit{i.e.} for
quantum field problems, the treatment of the theory on a complex contour is
hard to follow, because it is possible to have complicated Jacobian factors
\cite{bendx4}.

To avoid the existence of complicated Jacobian factors within the study of a
$\mathcal{PT}$-symmetric theory on a complex contour, one may seek a way to
modify the recipe used in quantum mechanical $\mathcal{PT}$-symmetric problems
in a manner that makes it applicable for $\mathcal{PT}$-symmetric field
theories. The usual recipe to study $\mathcal{PT}$-symmetric quantum theories
mentioned above has shown that the spectrum of the $\mathcal{PT}$-symmetric
$\left(  -x^{4}\right)  $ is bounded from below, although the classical
potential $\left(  -x^{4}\right)  $ is unstable. Therefore, the mentioned
algorithm can advocate the vacuum stability for the theory in $0+1$ space-time
dimensions. Before we go on, we need to assert that for $\mathcal{PT}%
$-symmetric quantum field theories, there exists a lack of studies in the
literature that discuss the vacuum stability for unstable classical potentials
like that of the $\mathcal{PT}$-symmetric $\left(  -\phi^{4}\right)  $ field
theory. In this work, we apply an algorithm which mimics the usual complex
contour method, and it avoids the problems associated with its direct
extension to quantum field problems. As we will show in this work, the
algorithm we use can explain the stability of the vacuum of the the
$\mathcal{PT}$-symmetric $\left(  -\phi^{4}\right)  $ field theory for which
classical analysis prohibits vacuum stability. In fact, the algorithm can be
applied to any quantum field theory but we use the $\mathcal{PT}$-symmetric
$\left(  -\phi^{4}\right)  $ theory as an illustrative example.

The algorithm we follow to study a $\mathcal{PT}$-symmetric \ field theory is
in the same spirit as the known complex contour method applied to the quantum
mechanical cases \cite{bendr,qop,bendrcop,jonespath,jonesqop}. In this
algorithm, we follow the canonical quantization method in which the
Hamiltonian determines the dynamics of the system \cite{Peskin}. Therefore,
the non-Hermiticity of the theory can be realized. Moreover, the canonical
quantization method employs two important features; (i) the equal time
canonical commutation relations and (ii) the Heisenberg picture for the
operators which leads the field to verify the Heisenberg equation of motion.
These features let the amplitudes obtained through this algorithm to know
about the metric \cite{green, jonesgr2}. Accordingly, the algorithm we use
avoids the calculation of the metric operator, which is hard to get for the
$\mathcal{PT}$-symmetric $\left(  -\phi^{4}\right)  $ field theory
\cite{jonesgr2}.

To account for the complex contour in the method we apply, we shift the field
$\phi$ to $\psi+B$, where $B$ is a $C$-number representing the vacuum
condensate. The field $\psi$ is real and has a different mass, while the
condensate is to be determined from the effective potential by constraining it
to satisfy the following stability conditions;
\begin{equation}
\frac{\partial V_{eff}}{\partial B}=0,\text{ \  \  \  \  \  \ }\frac{\partial
^{2}V_{eff}}{\partial B^{2}}=M^{2}, \label{rencon}%
\end{equation}
where $M$ is the renormalized mass of the field $\psi$. Since the renormalized
mass is always chosen to be real and positive, the effective potential as a
function of the condensate $B$ is bounded from below. However, as we will see
later, in this case $B$ ought to be imaginary, and thus the contour $\psi+B$
is complex. Hence, the resulting effective theory is non-Hermitian but
$\mathcal{PT}$-symmetric, which secures the reality of spectrum.

The conditions in Eq.(\ref{rencon}) guarantee a bounded-from-below effective
potential, and also agree with the known constraints applied to the effective
potential \cite{Peskin,Ryder}. In fact, the condition $\frac{\partial V_{eff}%
}{\partial B}=0$ is used to kill tad pole diagrams \cite{Peskin}, while
$\frac{\partial^{2}V_{eff}}{\partial B^{2}}=M^{2}$ represents the mass
renormalization condition \cite{Ryder}. To give an idea about how this
algorithm mimics the famous complex contour method, we mention that in quantum
mechanical studies we used to have localized wave functions ($\chi
\rightarrow0$ as $x\rightarrow \infty$) associated with bounded-from-below
potentials. Apparently, the conditions $\frac{\partial V_{eff}}{\partial B}=0$
and \  \  \ $\frac{\partial^{2}V_{eff}}{\partial B^{2}}=M^{2}$ define a minimum
in the effective potential. Therefore, the algorithm mimics the quantization
condition $\chi \rightarrow0$ as $\left \vert x\right \vert \rightarrow \infty$,
applied in the complex contour method. Within this regime, the field shift
$\phi \rightarrow \psi+B$ with $B$ imaginary resembles the choice of a complex
contour. For some theories, the spectrum is sensitive to the boundary
condition $\chi \rightarrow0$ as $\left \vert x\right \vert \rightarrow \infty$,
and thus the theory has different spectra for different contours. In this
case, in our algorithm, the conditions;
\begin{equation}
\frac{\partial V_{eff}}{\partial B}=0,\text{ \  \  \  \  \  \ }\frac{\partial
^{2}V_{eff}}{\partial B^{2}}=M^{2},\nonumber
\end{equation}
lead to different $B$ solutions, and the theory will have different vacua
defined by different condensate solutions.

For a quantitative test for the algorithm mentioned above in the study of
$\mathcal{PT}$-symmetric problems, we refer to our previous work in
Ref.\cite{aboqm}. There, we applied the effective field algorithm for the
calculations within the quantum mechanical $\mathcal{PT}$-symmetric $\left(
-x^{4}\right)  $ theory. We found reasonable results for the energy spectrum
and the vacuum condensate compared to exact results. Also, we obtained the
relations;%
\begin{align}
B  &  =-\sqrt{\frac{M^{2}}{-4g}},\nonumber \\
M  &  =\sqrt[3]{6g}, \label{param}%
\end{align}
for the vacuum condensate $B$ and the effective mass of the massless
$\mathcal{PT}$-symmetric $\left(  -x^{4}\right)  $ theory. These relations
have been reproduced by Jones in Ref. \cite{jonesgr2} using the
Schwinger-Dyson equations \footnote{Take into account the relations between
our coupling and their coupling $\left(  \lambda=2g\right)  $ and a rescaling
$\frac{1}{2}$ to their Hamiltonian is to be taken into account}. Such kind of
interesting results support the extension of the algorithm to quantum field
theories (higher dimensions) which is our aim in this work. In fact, we will
tackle the point of vacuum stability of the $\mathcal{PT}$-symmetric $\left(
-\phi^{4}\right)  $ theory, which has not been stressed before in the
literature. However, since in higher dimensions there exist UV divergences in
the calculations, one has to employ known tools to cure them. For that, the
algorithm we apply starts by using a normal ordered theory. To eliminate
divergences at the first order in the coupling, one normal order the theory
with respect to another mass parameter. This technique has been used in the
context of super renormalizable quantum field theories in
Refs.\cite{coleman,chang2}.

The paper is organized as follows. In Section \ref{formu}, the formulation of
the effective field method is introduced. The calculation of the effective
potential up to $g^{1}$ and $g^{2}$ order of approximations for the
$\mathcal{PT}$-symmetric $\left(  -\phi^{4}\right)  $ field theory in $1+1$
space-time dimensions is presented in Section \ref{eff1p1-s}, while the $2+1$
case is considered in Section \ref{effe2p1-s}. In Section \ref{conc}, the
discussions and conclusions are introduced.

\section{Formulation of effective field method\label{formu}}

In the absence of an external source, the effective potential is equivalent to
the vacuum energy $E$ ( $E=\langle0|H|0\rangle$). To illustrate the
implementation of the above mentioned ideas for the calculation of the
effective potential of the $\mathcal{PT}$-symmetric $\left(  -\phi^{4}\right)
$ theory, we start by the Hamiltonian density of the form;
\begin{equation}
H=N_{m}\left(  \frac{1}{2}\left(  \left(  \nabla \phi \right)  ^{2}+\pi
^{2}+m^{2}\phi^{2}\right)  -\frac{g}{4}\phi^{4}\right)  ,\label{origham}%
\end{equation}
in which $N_{m}$ indicates that $H$ is a normal-ordered form with respect to
the vacuum of the field $\phi$ of mass $m$. In introducing the field shift
$\phi \rightarrow \psi+B$ , the Hamiltonian density takes the form;%
\[
H\rightarrow N_{m}\left(  \frac{1}{2}\left(  \left(  \nabla \psi \right)
^{2}+\pi^{2}+m^{2}\left(  \psi+B\right)  ^{2}\right)  -\frac{g}{4}\left(
\psi+B\right)  ^{4}\right)  .
\]
Also, in taking into account the relation \cite{coleman};
\begin{equation}
N_{m}\exp \left(  i\beta \psi \right)  =\exp \left(  -\frac{1}{2}\beta^{2}%
\Delta \right)  N_{M}\exp \left(  i\beta \psi \right)  \text{,}\label{normal2}%
\end{equation}
one can obtain the resulting Hamiltonian normal-ordered with respect to the
new mass parameter $M$ of the effective field $\psi$. To show this, we first
note that this relation can lead to the following set of relations;%

\begin{align}
N_{m}\psi &  =N_{M}\psi,\nonumber \\
N_{m}\psi^{2}  &  =N_{M}^{2}\psi^{2}+\Delta,\nonumber \\
N_{m}\psi^{3}  &  =N_{M}\psi^{3}+3\Delta N_{M}\psi,\label{normall}\\
N_{m}\psi^{4}  &  =N_{M}\psi^{4}+6\Delta N_{M}\psi^{2}+3\Delta^{2},\nonumber
\end{align}
where $\Delta$ is the free field two point function \cite{coleman}. For the
kinetic term, we can get the result;
\begin{equation}
N_{m}\left(  \frac{1}{2}\left(  \nabla \psi \right)  ^{2}+\frac{1}{2}\pi
^{2}\right)  =N_{M}\left(  \frac{1}{2}\left(  \nabla \psi \right)  ^{2}+\frac
{1}{2}\pi^{2}\right)  +E_{0}(M)-E_{0}(m),
\end{equation}

where%
\begin{align}
E_{o}(\Omega)  &  =\frac{1}{4}\int \frac{d^{D-1}k}{\left(  2\pi \right)  ^{D-1}%
}\left(  \frac{2k^{2}+\Omega^{2}}{\sqrt{k^{2}+\Omega^{2}}}\right)
,\nonumber \\
&  =\frac{1}{2}\frac{1}{\left(  4\pi \right)  ^{\frac{D-1}{2}}}\frac{D-1}%
{2}\left(  \frac{\Gamma \left(  \frac{1}{2}-\frac{D-1}{2}-1\right)  }%
{\Gamma \left(  \frac{1}{2}\right)  }\left(  \frac{1}{\Omega^{2}}\right)
^{\frac{1}{2}-\frac{D-1}{2}-1}\right) \nonumber \\
&  +\frac{\Omega^{2}}{4}\frac{1}{\left(  4\pi \right)  ^{\frac{D-1}{2}}}\left(
\frac{\Gamma \left(  \frac{1}{2}-\frac{D-1}{2}\right)  }{\Gamma \left(  \frac
{1}{2}\right)  }\left(  \frac{1}{\Omega^{2}}\right)  ^{\frac{1}{2}-\frac
{D-1}{2}}\right) \label{Eomega}\\
&  =\frac{1}{8}\left(  \frac{1}{2}\right)  ^{D}\Gamma \left(  -\frac{1}%
{2}D\right)  \left(  2D-4\right)  \Omega^{D}\pi^{-\frac{1}{2}D}.\nonumber
\end{align}
Here $D$ is the dimension of the space-time. Considering these forms, one can
rewrite the Hamiltonian density $H$ in Eq.(\ref{origham}) in the form;
\begin{align}
H  &  =N_{m}\left(  \frac{1}{2}\left(  \nabla \psi \right)  ^{2}+\frac{1}{2}%
\pi^{2}+\frac{1}{2}m^{2}\left(  \psi+B\right)  ^{2}-\frac{g}{4}\left(
\psi+B\right)  ^{4}\right) \nonumber \\
&  =N_{M}\left(
\begin{array}
[c]{c}%
\frac{1}{2}\left(  \nabla \psi \right)  ^{2}+\frac{1}{2}\pi^{2}+\left(  \frac
{1}{2}m^{2}-\frac{3}{2}B^{2}g\right)  \left(  \psi^{2}+\Delta \right)
-\frac{1}{4}g\left(  \psi^{4}+6\Delta \psi^{2}+3\Delta^{2}\right) \\
-Bg\left(  \psi^{3}+3\Delta \psi \right)  +\left(  Bm^{2}-B^{3}g\right)  \psi \\
+\left(  \frac{1}{2}B^{2}m^{2}-\frac{1}{4}B^{4}g\right)  +E_{0}(M)-E_{0}(m)
\end{array}
\right) \nonumber \\
&  =N_{M}\left(
\begin{array}
[c]{c}%
\frac{1}{2}\left(  \nabla \psi \right)  ^{2}+\frac{1}{2}\pi^{2}+\left(  \frac
{1}{2}m^{2}-\frac{3}{2}gB^{2}-\frac{3}{2}g\Delta \right)  \psi^{2}-Bg\psi
^{3}-\frac{1}{4}g\psi^{4}\\
+\left(  Bm^{2}-gB^{3}-3g\Delta B\right)  \psi \\
\Delta \left(  \frac{1}{2}m^{2}-\frac{3}{2}B^{2}g\right)  -\frac{3}{4}%
g\Delta^{2}\allowbreak+\left(  \frac{1}{2}B^{2}m^{2}-\frac{1}{4}B^{4}g\right)
+E_{o}(M)-E_{o}(m).
\end{array}
\right)  . \label{normalM}%
\end{align}
In fact, $\Delta$ and $E_{0}$ might be divergent in space-time dimensions
higher than one. The divergences can be eliminated as it was done by Coleman
in Ref. \cite{coleman}, where the propagator of mass $m$ is subtracted from
that of the mass $M$ of the effective field. In Ref. \cite{Mag}, this
regularization method has been used also to regularize the sunset diagram. So,
we shall use this regularization method even for contributions to the
effective potential beyond the normal ordering result.

The effective potential, or equivalently the vacuum energy can be obtained
from Eq.(\ref{normalM}) where normal-ordered fields result in zero vacuum
expectation values, and thus do not contribute to the effective potential. In
the formula above for the effective Hamiltonian, the quantities $\Delta$ and
$E_{o}$ depend on the dimension of the space-time. Accordingly, we will study
the $1+1$ and $2+1$ cases individually.

\section{The effective potential of the $\mathcal{PT}$-symmetric $\left(
-\phi^{4}\right)  _{1+1}$ field theory\label{eff1p1-s}}

The Hamiltonian form in Eq.(\ref{normalM}) includes the space-time dependent
terms $\Delta$ and $E_{o}$. In $1+1$ space-time dimensions, one can expand
$E_{o}(\Omega)$ in Eq.(\ref{Eomega}) as a power series in $\epsilon=D-2$ to
get the result;%
\begin{equation}
E_{0}(\Omega)=\frac{1}{8}\frac{\Omega^{2}}{\pi}+O\left(  \epsilon \right)  ,
\end{equation}
and thus, we obtain the following form;
\begin{equation}
E_{0}(M)-E_{0}(m)=\frac{1}{8\pi}\left(  M^{2}-m^{2}\right)  .
\end{equation}
This is exactly the result obtained in Ref.\cite{coleman}. The vacuum energy
is then given by;%
\begin{equation}
E=\langle0|H|0\rangle=\Delta \left(  \frac{1}{2}m^{2}-\frac{3}{2}B^{2}g\right)
-\frac{3}{4}g\Delta^{2}\allowbreak+\left(  \frac{1}{2}B^{2}m^{2}-\frac{1}%
{4}B^{4}g\right)  +\frac{1}{8\pi}\left(  M^{2}-m^{2}\right)  ,
\end{equation}
with $\Delta=-\frac{1}{4\pi}\ln t$ and $t=\frac{M^{2}}{m^{2}}$. This result
has been obtained relying on the fact that the vacuum expectation values of
the normal-ordered operators in Eq. (\ref{normalM}) are certainly zero, and we
are left with the field-independent terms (last line in Eq.(\ref{normalM})).
To cure the divergences that appear in the calculations of $\Delta$, we
subtracted the propagator with mass $m$ from that with $M$ \ as in Ref.
\cite{coleman}.

\begin{figure}[ptbh]
\centering
\includegraphics{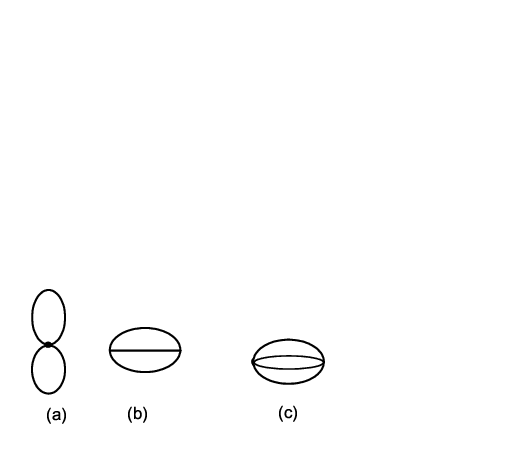} \caption{The Feynman diagrams ( up to second
order in the coupling) contributing to the vacuum energy of the $\mathcal{PT}%
$-symmetric $\left(  -\phi^{4}\right)  $ theory. Diagram (a) is a cactus
diagram for which normal-ordering accounts for its contribution to the vacuum
energy.}%
\label{Graph1}%
\end{figure}

The above result for the vacuum energy accounts for the contribution of the
one vertex Feynman diagram ( diagram (a) in Fig.\ref{Graph1}) to the effective
potential. In the absence of external source, the effective potential is
equivalent to the vacuum energy \cite{Peskin}, and it has to satisfy the
conditions \cite{Ryder};
\begin{align}
\frac{\partial E\left(  M,B,g\right)  }{\partial B}  &  =0,\nonumber \\
\frac{\partial^{2}E\left(  M,B,g\right)  }{\partial B^{2}}  &  =M^{2}.
\label{ren}%
\end{align}
In using the parameters redefinition; $b^{2}=4\pi B^{2}$,
$t=\frac{M^{2}}{m^{2}}$, and $G=\frac{g}{2\pi m^{2}}$, one gets;
\begin{equation}
e=\frac{8\pi E}{m^{2}}=b^{2}-G\left(  \frac{1}{4}b^{4}+\frac{3}{4}\ln
^{2}t-\frac{3}{2}b^{2}\ln t\right)  +\left(  t-\ln t-1\right)  .
\label{vac-en-nor}%
\end{equation}
The condition $\frac{\partial E}{\partial B}=0$ leads to the relation
\begin{equation}
\left(  -Gb^{2}+\left(  2+3G\ln t\right)  \right)  b=0,
\end{equation}
$\allowbreak$ where for $b\neq0$, it results in the solution $t=\exp \left(
\frac{1}{3}\frac{Gb^{2}-2}{G}\right)  $. In using the relation; $\frac
{\partial^{2}E\left(  M,B,g\right)  }{\partial B^{2}}=M^{2}$, one can show
that $b^{2}=\frac{-t}{G}$, but in this case both conditions in Eq.(\ref{ren})
are used which means that the obtained parameters ($b$ and $t$) define the
minimum of the effective potential. Equivalently, we get the result;
\begin{align}
b^{2}  &  =-\frac{t}{G}=-\frac{\exp \left(  \frac{1}{3}\frac{Gb^{2}-2}%
{G}\right)  }{G},\nonumber \\
b  &  =\sqrt{-3W\left(  \frac{1}{3G}e^{-\frac{2}{3G}}\right)  ,}%
\end{align}
where $W$ is the Lambert's $W$ function defined by $W(x)e^{W(x)}=x$. \ Note
that $W\left(  x\right)  =\allowbreak x+O\left(  x^{2}\right)  $, for small
values of the argument $x$. Therefore, we obtain the result;
\begin{equation}
b_{G\rightarrow0^{+}}=\pm i\frac{1}{\sqrt{G}}e^{-\frac{1}{3G}}. \label{bexp}%
\end{equation}
This exponential behavior for the dependence of the vacuum condensate on the
coupling has been obtained before in Ref. \cite{bendvac}, which constitutes a
good test for our calculations.

To advocate the vacuum stability, one can use the the relation $t=\exp \left(
\frac{1}{3}\frac{Gb^{2}-2}{G}\right)  $ to plot the vacuum energy in
Eq.(\ref{vac-en-nor}). As shown in Fig. \ref{normal-en}, the effective
potential is bounded from below, and thus the plot shows the stability of the
vacuum state. This result is pretty interesting, as it is the first time to
show that the vacuum of the $\mathcal{PT}$-symmetric $\left(  -\phi
^{4}\right)  $ scalar field theory is stable in $1+1$ space-time dimensions.

A note to be mentioned is that for imaginary $b$, the effective Hamiltonian
obtained in Eq.(\ref{normalM}) is non-Hermitian, but it is $\mathcal{PT}%
$-symmetric. Also, the $B\psi^{3}$ term turns the theory well defined on the
real line \cite{bendrcop}.

One can go beyond the above result for the vacuum energy and include the
radiative corrections received from the sunset ( diagram (b) in
Fig.\ref{Graph1}), and the watermelon ( diagram (c) in Fig.\ref{Graph1})
diagrams. These diagrams constitute the $G^{2}$ contribution to the effective
potential, which then takes the form;
\begin{equation}
\frac{8\pi E}{m^{2}}=b^{2}-G\left(  \frac{1}{4}b^{4}+\frac{3}{4}\ln^{2}%
t-\frac{3}{2}b^{2}\ln t\right)  +\left(  t-\ln t-1\right)  -G^{2}\left(
\alpha b^{2}\frac{1}{t}+\beta \left(  \frac{1}{t}-1\right)  \right)  ,
\end{equation}
with $\beta=3.155$ and $\alpha=\frac{1}{2}\left(  \Psi \left(  \frac{1}%
{3},1\right)  -\Psi \left(  \frac{2}{3},1\right)  \right)  , $ while
$\Psi \left(  x,n\right)  =\frac{d^{n+1}}{dx^{n+1}}\ln \Gamma \left(  x\right)  $
( see the appendix for the calculation of the Feynman diagrams).

In applying the condition $\frac{\partial E}{\partial B}=0,$ the coefficient
of $\psi$ is always zero, and thus the above result does not include Feynman
diagrams resulting from the $\psi$ term in the Hamiltonian in
Eq.(\ref{normalM}). Accordingly, the stability requirement for which one
always subject $E$ to the condition $\frac{\partial E}{\partial b}=0$ then
yields the result;
\begin{equation}
\left(  \left(  -G\right)  b^{2}+\frac{1}{t}\left(  3t\left(  \ln t\right)
G-2\alpha G^{2}+2t\right)  =0\right)  b=0.
\end{equation}
For $b\neq0$, one can solve for $t$ to have the form;%
\begin{equation}
t=\frac{\frac{2}{3}\alpha G}{W\left(  \frac{2}{3}\alpha Ge^{x}\right)  },
\end{equation}
where $x=\frac{2-Gb^{2}}{3G}$. Again, when we substitute this result in $E$ ,
and for $b$ imaginary, we get the bounded-from-below effective potential
plotted in Fig. \ref{e1p1}$.$

\section{The effective potential of the $\mathcal{PT}$-symmetric $\left(
-\phi^{4}\right)  _{2+1}$ field theory\label{effe2p1-s}}

For further confirmation of the stability of the vacuum of the $\mathcal{PT}%
$-symmetric $\left(  -\phi^{4}\right)  $ scalar field theory in other
space-time dimensions, we consider the $2+1$ dimensions case.

In this case, $\Delta$ in Eq.(\ref{normalM} ) takes the form;
\begin{align}
\Delta &  =\frac{1}{^{\left(  2\pi \right)  ^{3}}}\left(  \int \frac{d^{3}%
p}{p^{2}-M^{2}}-\int \frac{d^{3}k}{k^{2}-m^{2}}\right) \nonumber \\
&  =\frac{1}{\left(  4\pi \right)  ^{\frac{3}{2}}}\left(  \frac{\Gamma \left(
1-\frac{3}{2}\right)  }{\left(  M^{2}\right)  ^{1-\frac{3}{2}}}\right)
-\frac{1}{\left(  4\pi \right)  ^{\frac{3}{2}}}\left(  \frac{\Gamma \left(
1-\frac{3}{2}\right)  }{\left(  m^{2}\right)  ^{1-\frac{3}{2}}}\right)
\nonumber \\
&  =\frac{1}{4\pi}\left(  m-M\right)  ,
\end{align}
and $E_{0}(\Omega)$ in Eq.(\ref{Eomega}) is given by;
\begin{equation}
E_{o}(\Omega)=\frac{1}{24\pi}\Omega^{3}.
\end{equation}
After substituting for the values of $\Delta$ and $E_{0}$ in Eq.(\ref{normalM}%
), we get;%

\begin{equation}
E=\langle0|H|0\rangle=\Delta \left(  \frac{1}{2}m^{2}-\frac{3}{2}B^{2}g\right)
-\frac{3}{4}g\Delta^{2}\allowbreak+\left(  \frac{1}{2}B^{2}m^{2}-\frac{1}%
{4}B^{4}g\right)  +\frac{1}{24\pi}\left(  M^{3}-m^{3}\right)  ,\text{ }%
\end{equation}
or,
\begin{equation}
\frac{8\pi E}{m^{3}}=-\frac{1}{2}Gb^{4}+b^{2}\left(  3G\left(  t-1\right)
+1\right)  -\frac{3}{2}G\left(  t-1\right)  ^{2}+\frac{1}{3}\left(
t^{3}-3t+2\right)  ,
\end{equation}
where $G=\frac{g}{4\pi m}$, $t=\frac{M}{m}$, and $b=B\sqrt{\frac{4\pi}{m}}$.
Since the most important corrections to the effective potential come from
logarithmic contributions, one has to include at least the $G^{2}$ corrections
to obtain a reliable contribution to the effective potential. This leads to
the result;
\begin{align}
\frac{8\pi E}{m^{3}}  &  =-\frac{1}{2}Gb^{4}+b^{2}\left(  3G\left(
t-1\right)  +1\right)  -\frac{3}{2}G\left(  t-1\right)  ^{2}\nonumber \\
&  +6G^{2}b^{2}\ln t+\frac{1}{3}\left(  t^{3}-3t+2\right)  -9G^{2}\left(
t-1\right)  \ln t.
\end{align}
\bigskip Similar to the $1+1$ case, in applying the condition $\frac{\partial
E}{\partial b}=0$, we get;%
\begin{equation}
\left(  -2G\right)  b^{2}+\left(  12G^{2}\ln t+6G\left(  t-1\right)
+2\right)  b=0,
\end{equation}
$\allowbreak$and for $b\neq0$, we have
\begin{equation}
t=2GW\left(  \frac{1}{2G}e^{\frac{1}{6}\frac{Gb^{2}+3G-1}{G^{2}}}\right)  .
\end{equation}
Again, in substituting this result into the form of $E$, we obtain the
bounded- from-below effective potential shown in Fig.\ref{e2p1}. This is
correct as long as $b$ is kept imaginary. Also, one can follow the same
argument led to Eq.(\ref{bexp}) for the $1+1$ case to show that the $b$ value
at the minimum of the effective potential behaves like;
\begin{equation}
b_{G\rightarrow0^{+}}=\pm \frac{1}{2}i\frac{\sqrt{2}}{\sqrt{G}}e^{-\frac
{1}{12G}}. \label{bexp2p1f}%
\end{equation}
Such exponential behavior has also been obtained in Ref.\cite{bendvac}, which
constitutes a good check for the accuracy of our calculations.

\section{Discussions and Conclusions\label{conc}}

We employed the canonical quantization method for the calculation of the
effective potential for the $\mathcal{PT}$-symmetric $\left(  -\phi
^{4}\right)  $ scalar field theory. We considered the cases of the $1+1$ and $2+1$ space-time dimensions individually.
We have done that although in  the literature the effective potential is often studied within the path
integral formulation of the theory \cite{Peskin}. In fact, the path integral
formulation by itself is obtained via the canonical quantization of the
theory, for which the equal time canonical relations of the form;%
\begin{equation}
\left[  \phi \left(  x_{1},t\right)  ,\pi \left(  x_{2},t\right)  \right]
=i\delta^{D}\left(  x_{1}-x_{2}\right)  ,
\end{equation}
are satisfied. Our point in following the canonical quantization method is that, with in this regime, the
Hamiltonian operator determines the dynamics of the system. Therefore,  the non-Hermiticity of the Hamiltonian operator for a $\mathcal{PT}$-symmetric field theory can be realized easily. So, we find it more plausible  to work with
operators ( canonical quantization)  than working with integration over classical
functionals ( path integral).  Note that, in the
canonical quantization of a theory, one also employs the Heisenberg picture
for the operators ( see the chapters in the first part in Ref.\cite{Peskin}).
Thus, for the theory under consideration, the Heisenberg equation of motion is
satisfied. Accordingly, the calculated amplitudes know about the metric
\cite{green,jonesgr2}. This means that the followed algorithm in our work
avoids the calculation of the metric operator, which is hard to get for the
theory under consideration.

For the $\mathcal{PT}$-symmetric $\left(  -\phi^{4}\right)  $ scalar field
theory, the classical potential is bounded from above. Consequently, the common classical
analysis  predicts an unstable  vacuum. In our work, we have shown that the effective potential is bounded-from-below which shows that the vacuum state of the $\mathcal{PT}$-symmetric $\left(  -\phi ^{4}\right)  $ scalar field theory is stable. This result tells us that classical analysis are not always reliable either quantitatively or qualitatively.  

The stability of the theory is constrained by the existence of an imaginary condensate. The imaginary value of the condensate
renders the effective theory non-Hermitian but $\mathcal{PT}$-symmetric. In
fact, the effective theory \ is well defined on the real line because of the
existence of the pure imaginary, $B\psi^{3}$, term in the Hamiltonian.

To test the accuracy of our results, we obtained the vacuum condensate at the
minimum of the effective potential. The behavior of the condensate has been found to approach its zero value for small coupling in an exponential manner
( Eq.(\ref{bexp})\&Eq.(\ref{bexp2p1f})). This exponential behavior has been
obtained before in Ref.\cite{bendvac}, which represents a good test for the
accuracy of our results.

This work sheds light on some how a new strange behavior of the quantum world.
It tells us that classical analysis does not always rule the quantum behavior
of a quantum particle. The situation is very similar to the tunneling effect
in quantum physics for which classical analysis totally prohibits tunneling
from existence, while the quantum world admits it. Likewise, the vacuum
stability is totally prohibited from a classical point of view for bounded-
from-above potentials, while we have shown that the potential felt by the
quantum particle is bounded from below, and thus allows a stable vacuum.

\begin{acknowledgments}
We would like to thank M. Al-Hashimi for his help in revising the manuscript.
\end{acknowledgments}

\begin{appendices}
\begingroup
\numberwithin{equation}{subsection}
\begin{center}
\appendix{\textbf{Appendix: Feynman Diagram Calculations}}
\end{center}
\subsection{The Sunset Diagram}
The sunset diagram (diagram (b) in Fig. .\ref{Graph1}) involves the integral;%
\begin{equation}
\label{is}I_{s}=\int \frac{d^{D}q}{\left(  2\pi \right)  ^{D}}\int \frac{d^{D}%
w}{\left(  2\pi \right)  ^{D}}\frac{1}{\left(  q^{2}-m^{2}\right)  \left(
w^{2}-m^{2}\right)  \left(  \left(  q+w\right)  ^{2}-m^{2}\right)  },
\end{equation}
In introducing the Feynman parameters $x,y$ and $z$ \cite{Peskin}, we get the following result;
\begin{align}
&  \frac{1}{\left(  q^{2}-m^{2}\right)  \left(  w^{2}-m^{2}\right)  \left(
\left(  q+w\right)  ^{2}-m^{2}\right)  }\\
&  =\int dx\int dy\int dz\text{ }\delta \left(  1-x-y-z\right)  \frac{\left(
n-1\right)  !}{\left(  x\left(  q^{2}-m^{2}\right)  +y\left(  w^{2}%
-m^{2}\right)  +z\left(  \left(  q+w\right)  ^{2}-m^{2}\right)  \right)  ^{n}},
\end{align}
where $n=3$.  Also, we can obtain the following result,%
\begin{align}
&  \left(  x\left(  q^{2}-m^{2}\right)  +y\left(  w^{2}-m^{2}\right)
+z\left(  \left(  q+w\right)  ^{2}-m^{2}\right)  \right)  ^{n}\nonumber \\
&  =\left(  x+z\right)  ^{n}\left(  q^{2}+\frac{2wzq}{\left(  x+z\right)
}-\frac{\left(  m^{2}x+y\left(  m^{2}-w^{2}\right)  +z\left(  m^{2}%
-w^{2}\right)  \right)  }{\left(  x+z\right)  }\right)  ^{n}\nonumber \\
&  =\left(  x+z\right)  ^{n}\left(  \left(  q+\frac{wzq}{\left(  x+z\right)
}\right)  ^{2}-\left(  \frac{wzq}{\left(  x+z\right)  }\right)  ^{2}%
-\frac{\left(  m^{2}x+y\left(  m^{2}-w^{2}\right)  +z\left(  m^{2}%
-w^{2}\right)  \right)  }{\left(  x+z\right)  }\right)  ^{n}\nonumber \\
&  =\left(  x+z\right)  ^{n}\left(  q^{2}-\left(  \frac{wz}{\left(
x+z\right)  }\right)  ^{2}-\frac{\left(  m^{2}x+y\left(  m^{2}-w^{2}\right)
+z\left(  m^{2}-w^{2}\right)  \right)  }{\left(  x+z\right)  }\right)  ^{n}.
\end{align}
In using the Euclidean variable $q_{E}$, such that $q_{E}^{0}=-iq^{0}$, and
$q_{E}^{i}=q^{i}$, the integral over the internal momentum $q$ will take the form;%
\begin{equation}
I_{q}=\int \frac{d^{D}q_{E}}{\left(  2\pi \right)  ^{D}}\frac{id^{D}q_{E}%
}{\left(  x+z\right)  ^{n}\left(  -1\right)  ^{n}\left(  q_{E}^{2}+\left(
\frac{wz}{\left(  x+z\right)  }\right)  ^{2}+\frac{\left(  m^{2}x+y\left(
m^{2}-w^{2}\right)  +z\left(  m^{2}-w^{2}\right)  \right)  }{\left(
x+z\right)  }\right)  ^{n}}.
\end{equation}
The result of the $q$-integration  is then;
\begin{equation}
I_{q}=\frac{\left(  n-1\right)  !i}{\left(  x+z\right)  ^{n}\left(  -1\right)
^{n}}\frac{1}{\left(  4\pi \right)  ^{\frac{D}{2}}}\frac{\Gamma \left(
n-\frac{D}{2}\right)  }{\Gamma \left(  n\right)  }\frac{1}{\left(  \left(
\frac{z^{2}}{\left(  x+z\right)  ^{2}}-\frac{1}{x+z}\left(  y+z\right)
\right)  w^{2}+\frac{1}{x+z}\left(  m^{2}x+m^{2}y+m^{2}z\right)
\allowbreak \right)  ^{n-\frac{D}{2}}}.
\end{equation}
Similarly, the integration over the internal momentum $w$ can be obtained as;
\begin{align}
I_{w}  &  =\int \frac{d^{D}w}{\left(  \frac{z^{2}}{\left(  x+z\right)  ^{2}%
}-\frac{1}{x+z}\left(  y+z\right)  \right)  ^{n-\frac{D}{2}}\left(
w^{2}+\frac{1}{x+z}\frac{\left(  m^{2}x+m^{2}y+m^{2}z\right)  }{\left(
\frac{z^{2}}{\left(  x+z\right)  ^{2}}-\frac{1}{x+z}\left(  y+z\right)
\right)  }\allowbreak \right)  ^{n-\frac{D}{2}}}\nonumber \\
&  =\int \frac{id^{D}w_{E}}{\left(  \frac{z^{2}}{\left(  x+z\right)  ^{2}%
}-\frac{1}{x+z}\left(  y+z\right)  \right)  ^{n-\frac{D}{2}}\left(  -1\right)
^{n-\frac{D}{2}}\left(  w_{E}^{2}-\frac{1}{x+z}\frac{\left(  m^{2}%
x+m^{2}y+m^{2}z\right)  }{\left(  \frac{z^{2}}{\left(  x+z\right)  ^{2}}%
-\frac{1}{x+z}\left(  y+z\right)  \right)  }\allowbreak \right)  ^{n-\frac
{D}{2}}}\nonumber \\
&  =\int \frac{id^{D}w_{E}}{\left(  \frac{z^{2}}{\left(  x+z\right)  ^{2}%
}-\frac{1}{x+z}\left(  y+z\right)  \right)  ^{n-\frac{D}{2}}\left(  -1\right)
^{n-\frac{D}{2}}\left(  w_{E}^{2}-\frac{1}{x+z}\frac{\left(  m^{2}%
x+m^{2}y+m^{2}z\right)  }{\left(  \frac{z^{2}}{\left(  x+z\right)  ^{2}}%
-\frac{1}{x+z}\left(  y+z\right)  \right)  }\allowbreak \right)  ^{n-\frac
{D}{2}}}\nonumber \\
&  =\frac{1}{\left(  4\pi \right)  ^{\frac{D}{2}}}\frac{i\Gamma \left(
n-D\right)  }{\Gamma \left(  n-\frac{d}{2}\right)  \left(  -1\right)
^{n-\frac{D}{2}}\left(  \frac{z^{2}}{\left(  x+z\right)  ^{2}}-\frac{1}%
{x+z}\left(  y+z\right)  \right)  ^{n-\frac{D}{2}}\left(  -\frac{1}{x+z}%
\frac{\left(  m^{2}x+m^{2}y+m^{2}z\right)  }{\left(  \frac{z^{2}}{\left(
x+z\right)  ^{2}}-\frac{1}{x+z}\left(  y+z\right)  \right)  }\allowbreak
\right)  ^{n-D}}.%
\end{align}
Therefore, $I_{s}$ in Eq.(\ref{is}) takes the form;
\begin{align}
I_{s} =-m^{6-2D}F_{\Gamma}\int_{0}^{1}dx\int_{0}^{1-x}dy\left(  -x-y+x^{2}%
+xy+y^{2}\right)  ^{\frac{-D}{2}},
\end{align}
where%
\begin{equation}
F_{\Gamma}=\frac{\left(  n-1\right)  !i}{\left(  x+z\right)  ^{n}\left(
-1\right)  ^{n}}\frac{1}{\left(  4\pi \right)  ^{\frac{D}{2}}}\frac
{\Gamma \left(  n-\frac{D}{2}\right)  }{\Gamma \left(  n\right)  }\frac
{1}{\left(  4\pi \right)  ^{\frac{D}{2}}}\frac{i\Gamma \left(  n-D\right)
}{\Gamma \left(  n-\frac{D}{2}\right)  \left(  -1\right)  ^{n-\frac{D}{2}}}.
\end{equation}
The integrand \ $I_{xy}$ below can be simplified as;
\begin{align}
I_{xy}  &  =\frac{1}{\left(  -1\right)  ^{n}\left(  m^{2}\right)  ^{n-D}}%
\int_{0}^{1}\int_{0}^{1-x}\frac{dxdy}{\left(  x^{2}+xy-x+y^{2}-y\right)
^{\frac{D}{2}}}\nonumber \\
&  =\frac{1}{\left(  -1\right)  ^{n}\left(  m^{2}\right)  ^{n-D}}\left(
-2\right)  \int_{0}^{\frac{1}{2}}\int_{-\alpha}^{\alpha}\frac{d\alpha d\beta
}{\left(  3\alpha^{2}+\beta^{2}-2\alpha \right)  ^{D}}.
\end{align}
In $1+1$ dimensions;%
\begin{align}
I_{xy}  &  =\frac{1}{\left(  -1\right)  ^{n}\left(  m^{2}\right)  ^{n-D}%
}\left(  -2\right)  \int_{0}^{\frac{1}{2}}\int_{-\alpha}^{\alpha}\frac{d\alpha
d\beta}{\left(  3\alpha^{2}+\beta^{2}-2\alpha \right)  ^{\frac{1}{2}d}%
}\nonumber \\
&  =\frac{1}{\left(  -1\right)  ^{n}\left(  m^{2}\right)  ^{n-D}}\left(
-2\right)  \left(  -\frac{\Psi \left(  \frac{1}{3},1\right)
-\operatorname{\Psi}\left(  \frac{2}{3},1\right)  }{6}\right)  ,
\end{align}
where $\Psi \left(  x,m\right)  $ is the polygamma function given
by;
\begin{equation}
\operatorname{\Psi}\left(  x,m\right)  =\frac{d^{m+1}}{dx^{m+1}}\ln
\Gamma \left(  x\right)  ,
\end{equation}
Accordingly, the diagram contribution ($\Delta E_{s}$) to the vacuum energy is;%
\begin{align}
\frac{8 \pi t}{m^2}\Delta E_{s}  &  =\frac{8\pi \left(  i^{3}\right)  }{-i\left(  3!\times
2\right)  }\left(  \left(  -i\right)  2\pi3!G\right)  ^{2}\frac{b^{2}}{4\pi
}\left(  n-1\right)  !\frac{2}{\Gamma \left(  n\right)  }\nonumber \\
&  \times \Gamma \left(  n-D\right)  \left(  -1\right)  ^{\frac{1}{2}D}\left(
4\pi \right)  ^{-D}\left(  -\frac{\operatorname{Psi}\left(  \frac{1}%
{3},1\right)  -\operatorname{Psi}\left(  \frac{2}{3},1\right)  }{6}\right) \\
&  =-3.\, \allowbreak515\,9G^{2}b^{2}\nonumber,
\end{align}
where we divided by a symmetry factor of $3!\times2$.

In $2+1$ dimensions, the integral $I_{xy}$ can also be calculated, and we get;
\begin{align}
I_{xy}  &  =\frac{1}{\left(  -1\right)  ^{n}\left(  m^{2}\right)  ^{n-D}%
}\left(  -2\right)  \int_{0}^{\frac{1}{2}}\int_{-\alpha}^{\alpha}\frac{d\alpha
d\beta}{\left(  3\alpha^{2}+\beta^{2}-2\alpha \right)  ^{\frac{1}{2}d}}\nonumber \\
&  =\frac{1}{\left(  -1\right)  ^{n}\left(  m^{2}\right)  ^{n-D}}\left(
-2\right)  \left(  i\pi \right),
\end{align}
and thus;%
\begin{align}
\frac{8 \pi }{m^3}\Delta E_{s}  &  =\frac{8\pi \left(  i\right)  ^{3}}{-i\left(  3!\times
2\right)  }\left(  3!\left(  -2\pi iG\right)  \right)  ^{2}\frac{b^{2}}{4\pi
}\frac{i}{\left(  -1\right)  ^{n}}\frac{1}{\left(  4\pi \right)  ^{\frac{D}{2}%
}}\frac{1}{\left(  4\pi \right)  ^{\frac{D}{2}}}\frac{i}{\left(  -1\right)
^{n-\frac{D}{2}}}\frac{1}{\left(  -1\right)  ^{-n}}\nonumber \\
&  \times \left(  -1\right)  ^{n}\left(  -2\right)  \left(  i\pi \right)
\Gamma \left(  n-D\right)  \left(  \left(  M^{2}\right)  ^{D-n}-\left(
m^{2}\right)  ^{D-n}\right) \nonumber \\
&  =\left(
\begin{array}
[c]{c}%
\frac{8\pi \left(  i\right)  ^{3}}{-i\left(  3!\times2\right)  }\left(
3!\left(  -4\pi iG\right)  \right)  ^{2}\frac{b^{2}}{4\pi}\frac{i}{\left(
-1\right)  ^{n}}\frac{1}{\left(  4\pi \right)  ^{\frac{D}{2}}}\frac{1}{\left(
4\pi \right)  ^{\frac{D}{2}}}\\
\frac{i}{\left(  -1\right)  ^{n-\frac{D}{2}}}\frac{1}{\left(  -1\right)
^{-n}}\left(  -1\right)  ^{n}\left(  -2\right)  \left(  i\pi \right)
\end{array}
\right)  \left(  -2\ln t\right) \nonumber \\
&  =6G^{2}b^{2}\ln t.
\end{align}
In the above result, we used the power series expansion for the Gamma function
as;
\begin{equation}
\left(  m^{2}\right)  ^{\epsilon}\Gamma \left(  -\epsilon \right)
=-\epsilon^{-1}+\left(  -\gamma-\ln \frac{m^{2}}{\mu^{2}}\right)  +O\left(
\epsilon \right)  \allowbreak,
\end{equation}
$\allowbreak$where $\epsilon=D-3$, and $\gamma$ is the Euler number.
\subsection{ The Watermelon Diagram}
For diagram (c) in Fig. \ref{Graph1}, one can follow the same steps used in
the sunset diagram above to calculate its contribution. For this case, consider the
integral;%
\begin{align}
I_{W} &  =\int \frac{d^{D}p}{\left(  2\pi \right)  ^{D}}\int \frac{d^{D}%
q}{\left(  2\pi \right)  ^{D}}\int \frac{d^{D}w}{\left(  2\pi \right)  ^{D}%
}\nonumber \\
&  \times \frac{1}{\left(  p^{2}-m^{2}\right)  \left(  q^{2}-m^{2}\right)
\left(  w^{2}-m^{2}\right)  \left(  \left(  p+q+w\right)  ^{2}-m^{2}\right)
}.
\end{align}
After introducing the Feynman parameters and the Euclidean variables $q_{E}$
such that $q_{E}^{0}=-iq^{0}$ and $q_{E}^{i}=q^{i}$, one can get the result;
\begin{align}
I_{W} &  =F\left(  \Gamma \right)  \int dx\int dy\int dz\delta \left(
1-x-y-z-u\right)  \nonumber \\
&  \times \frac{1}{\left(  -1\right)  ^{n-\frac{D}{2}}\left(
uxy+uxz+uyz+xyz\right)  ^{\frac{D}{2}}m^{2n-3D}\allowbreak}\nonumber \\
&  =\frac{F\left(  \Gamma \right)  }{m^{2n-3D}\allowbreak \left(  -1\right)
^{n-\frac{D}{2}}}\int_{0}^{1}dx\int_{0}^{1-x}dy\int_{0}^{1-x-y}dzf\left(
x,y,z\right)  ,\\
f\left(  x,y,z\right)   &  =\frac{1}{\left(  xy-x^{2}y-xy^{2}-2xyz+xz-x^{2}%
z-xz^{2}+yz-y^{2}z-yz^{2}\right)  ^{\frac{D}{2}}\allowbreak},\nonumber
\end{align}
where $n=4$, and
\begin{align}
F\left(  \Gamma \right)   &  =\frac{\left(  n-1\right)  !i}{\left(  -1\right)
^{n}}\frac{1}{\left(  4\pi \right)  ^{\frac{d}{2}}}\frac{\Gamma \left(
n-\frac{D}{2}\right)  }{\Gamma \left(  n\right)  }\frac{i}{\left(  -1\right)
^{n-\frac{d}{2}}}\frac{1}{\left(  4\pi \right)  ^{\frac{d}{2}}}\frac
{\Gamma \left(  n-D\right)  }{\Gamma \left(  n-\frac{d}{2}\right)  }\frac
{i}{\left(  -1\right)  ^{n-d}}\frac{1}{\left(  4\pi \right)  ^{\frac{d}{2}}%
}\frac{\Gamma \left(  n-\frac{3D}{2}\right)  }{\Gamma \left(  n-D\right)
}\nonumber \\
&  =-\left(  -1\right)  ^{-2n+\frac{3D}{2}}8^{-D}\frac{\pi}{\Gamma \left(
n\right)  }\Gamma \left(  n-\frac{3D}{2}\right)  \frac{\left(  n-1\right)
!i}{\left(  -1\right)  ^{n}}\nonumber \\
&  =-\left(  -1\right)  ^{\frac{1}{2}-3n+\frac{3D}{2}}\left(  4\pi \right)
\Gamma \left(  n-\frac{3D}{2}\right)  .
\end{align}
In $1+1$ space-time dimensions, one can calculate the integral;
\begin{align}
I_{xyz}  & =\int_{0}^{1}dx\int_{0}^{1-x}dy\int_{0}^{1-x-y}dz\times \nonumber \\
& \frac{1}{\left(  xy-x^{2}y-xy^{2}-2xyz+xz-x^{2}z-xz^{2}+yz-y^{2}%
z-yz^{2}\right)  ^{\frac{D}{2}}\allowbreak},
\end{align}
numerically and get the diagram contribution ($\Delta E_{w}$ ) to the effective potential as;%
\begin{equation}
\frac{8 \pi }{m^2}\Delta E_{w}=-3.155G^{2}\left(  \frac{1}{t}-1\right)
\end{equation}
In $2+1$ space-time dimensions, although the diagram is finite from the
dimensional analysis point of view, it does have a sub divergent diagram (
diagram (b)) and one has to be careful in dealing with such diagram
calculations. This diagram has been calculated in Ref. \cite{Feyn-diag} but in
following the same regularization technique we used before ( subtracting the
diagram with mass $m$ from that with mass $M$) we get,%
\begin{align}
Diagram\text{ }(c)  &  =\frac{4}{\left(  4\pi \right)  ^{3}}m\left(  -\frac
{1}{4\epsilon}+2+\frac{1}{2}\ln2t+\ln4t\right) \nonumber \\
&  =\frac{4}{\left(  4\pi \right)  ^{3}}m\left(  -\frac{1}{4\epsilon}%
+2+\frac{5}{2}\ln2+\frac{3}{2}\ln t\right) \nonumber \\
&  \rightarrow \frac{4}{\left(  4\pi \right)  ^{3}}\left(  t-1\right)  \frac
{3}{2}\ln t\nonumber \\
&  =\frac{6}{\left(  4\pi \right)  ^{3}}\left(  t-1\right)  \ln t,
\end{align}
or
\begin{align}
\Delta E_{w}  &  =\frac{8\pi \left(  i\right)  ^{4}\left(  i\right)  ^{3}%
}{-i\left(  4!\times2\right)  }\left(  3!\left(  -4\pi iG\right)  \right)
^{2}\frac{6}{\left(  4\pi \right)  ^{3}}\left(  t-1\right)  \ln t\nonumber \\
&  =-9G^{2}\left(  t-1\right)  \ln t.
\end{align}
Note that, we used the fact that the renormalization scheme should be fixed \cite{collinbook}, which
means that $\frac{M}{\nu}=\frac{m}{\mu}=t$, where $\nu$ and $\mu$ are of mass
units introduced to have dimensionless logarithms.
\endgroup
\end{appendices}

\newpage

\begin{figure}[ptbh]
\centering
\includegraphics{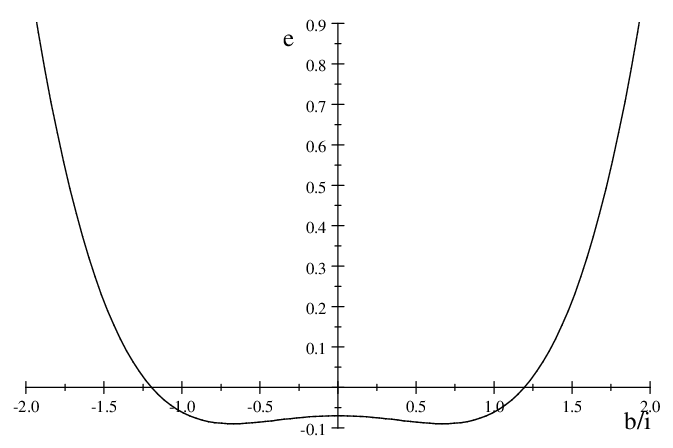} .
\caption{The effective potential
$e=\frac{8\pi E}{m^{2}}$, up to order $G^{1}$, versus the vacuum condensate
$b$ for $G=\frac{1}{2}$ for the $\mathcal{PT}$-symmetric$\left(  -\phi
^{4}\right)  $ scalar field theory in $1+1$ space-time dimensions.}%
\label{normal-en}%
\end{figure}

\newpage \begin{figure}[ptbh]
\begin{center}
\includegraphics{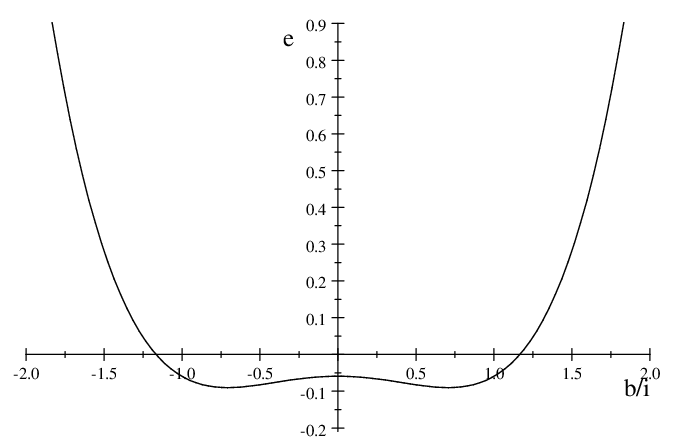}
\end{center}
\par
.\caption{The effective potential $e=\frac{8\pi E}{m^{2}}$ versus the vacuum
condensate $b$ for $G=\frac{1}{2}$ for the $\mathcal{PT}$-symmetric$\left(
-\phi^{4}\right)  $ scalar field theory in $1+1$ space-time dimensions, and  up to
$G^{2}$ order in the coupling.}%
\label{e1p1}%
\end{figure}

\begin{figure}[ptb]
\begin{center}
\includegraphics{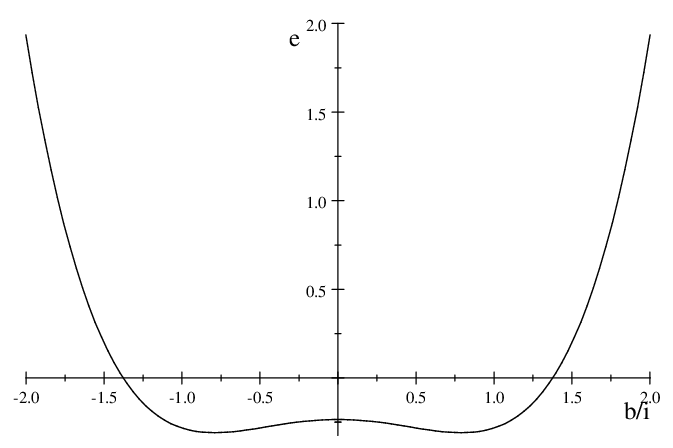}
\end{center}
\caption{The effective potential $e=\frac{8\pi E}{m^{3}}$, up to order $G^{2}%
$, versus the vacuum condensate $b$ for $G=\frac{1}{2}$ for the $\mathcal{PT}%
$-symmetric$\left(  -\phi^{4}\right)  $ scalar field theory in $2+1$
space-time dimensions.}%
\label{e2p1}%
\end{figure}


\newpage

\begin{thebibliography}{99}                                                                                               %


\bibitem {bendr}Carl Bender and Stefan Boettcher, Phys.Rev.Lett.80:5243-5246 (1998).

\bibitem {qop}H. F. Jones and R. J. Rivers, Phys.Rev.D75:025023 (2007).

\bibitem {bendrcop}Carl M. Bender, Dorje C. Brody and Hugh F. Jones, Phys.Rev.D73:025002(2006).

\bibitem {jonespath}H. F.Jones, J. Mateo and R. J. Rivers Phys.Rev.D74:125022 (2006).

\bibitem {jonesqop}H. F. Jones, J. Mateo, Phys.Rev.D73:085002 (2006).

\bibitem {aboqm}Abouzeid M.Shalaby, Phys. Rev. D 79, 065017 (2009).

\bibitem {bendx4}Carl M. Bender, Dorje C. Brody, Jun-Hua Chen, Hugh F. Jones,
Kimball A. Milton, and Michael C. Ogilvie1, Phys.Rev. D 74, 025016 (2006).

\bibitem {bendvac}Carl M. Bender, Peter N. Meisinger, and Haitang Yang,
Phys.Rev. D63, 045001 (2001).

\bibitem {Peskin}Book by Michael E.Peskin and Daniel V.Schroeder, "AN
INTRODUCTION TO THE QUANTUM FIELD THEORY" (1995).

\bibitem {green}H. F. Jones and R. J. Rivers, Phys. Lett. A 373, 3304 (2009).

\bibitem {jonesgr2}H.F. Jones, Int J Theor Phys 50: 1071--1080 (2011) .

\bibitem {Ryder}LEWIS H. RYDER , Quantum Field Theory (Second edition ,
CAMBRIDGE UNIVERSITY PRESS ) ( 1996).

\bibitem {coleman}Sideny Coleman, Phys.Rev.D11:2088 (1975).

\bibitem {chang2}S.~J.~Chang,
%``The Existence Of A Second Order Phase Transition In The Two-Dimensional Phi**4 Field Theory,''
Phys.\ Rev.\ D \textbf{13}, 2778 (1976) [Erratum-ibid.\ D \textbf{16}, 1979
(1976)].
%%CITATION = PHRVA,D13,2778;%%


\bibitem {Mag}Steven F. Magruder, Phys.\ Rev.\ D \textbf{14}, 1602 (1976).

\bibitem {quart}Jing-Ling Chen, L.C. Kwek and C.H.Oh, Phys. Rev. A 67, 012101 (2003).

\bibitem {bendx4q}Carl M. Bender , Dorje C. Brody and Hugh F. Jones,
Phys.Rev.D73:025002 (2006 ).

\bibitem {collinbook}John C. Collins, RENORMALIZATION, CAMBRIDGE UNIVERSITY
PRESS (1984).

\bibitem {Feyn-diag}Arttu K. Rajantie, Nucl.Phys. B480 ,729-752 ((1996));
Erratum-ibid. B513, 761-762 (1998).
\end{thebibliography}
\end{document}